\documentclass[prl,showpacs,superscriptaddress,floatfix]{revtex4} %preprint
\usepackage{graphicx}% Include figure files 
\usepackage{bm}% bold math   
\usepackage{epsfig} 
\usepackage{amsmath} 
\begin{document}

\title{Controlled Irradiative Formation of Penitentes} 

\author{Vance Bergeron} \affiliation{Ecole Normale Sup\'{e}reiure,
  Laboratoire de Physique, UMR 5672, 46 all\'{e}e d'Italie, 69364 Lyon
  Cedex 07, France} 

\author{Charles Berger} \affiliation{Ecole Normale Sup\'{e}reiure,
  Laboratoire de Physique Statistique, 24 rue Lhomond,75231 Paris
  Cedex 07, France}

\author{M. D. Betterton} \affiliation{Department of Physics, 390 UCB, University of Colorado, Boulder CO 80309, USA}
%\email{mdb@colorado.edu}
%\homepage{http://amath.colorado.edu/faculty/mdb}

\begin{abstract}
  Spike-shaped structures are produced by light-driven ablation in
  very different contexts. Penitentes 1-4 m high are common on Andean
  glaciers, where their formation changes glacier dynamics and
  hydrology. Laser ablation can produce cones 10-100 $\mu$m high with
  a variety of proposed applications in materials science. We report
  the first laboratory generation of centimeter-scale snow and ice
  penitentes. Systematically varying conditions allows identification
  of the essential parameters controlling the formation of ablation
  structures. We demonstrate that penitente initiation and coarsening
  requires cold temperatures, so that ablation leads to sublimation
  rather than melting. Once penitentes have formed, further growth of
  height can occur by melting. The penitentes intially appear as small
  structures (3 mm high) and grow by coarsening to 1-5 cm high. Our
  results are an important step towards understanding and controlling
  ablation morphologies.

\end{abstract}
\maketitle 

\begin{figure*}[t]
  \centering \includegraphics[height=5cm]{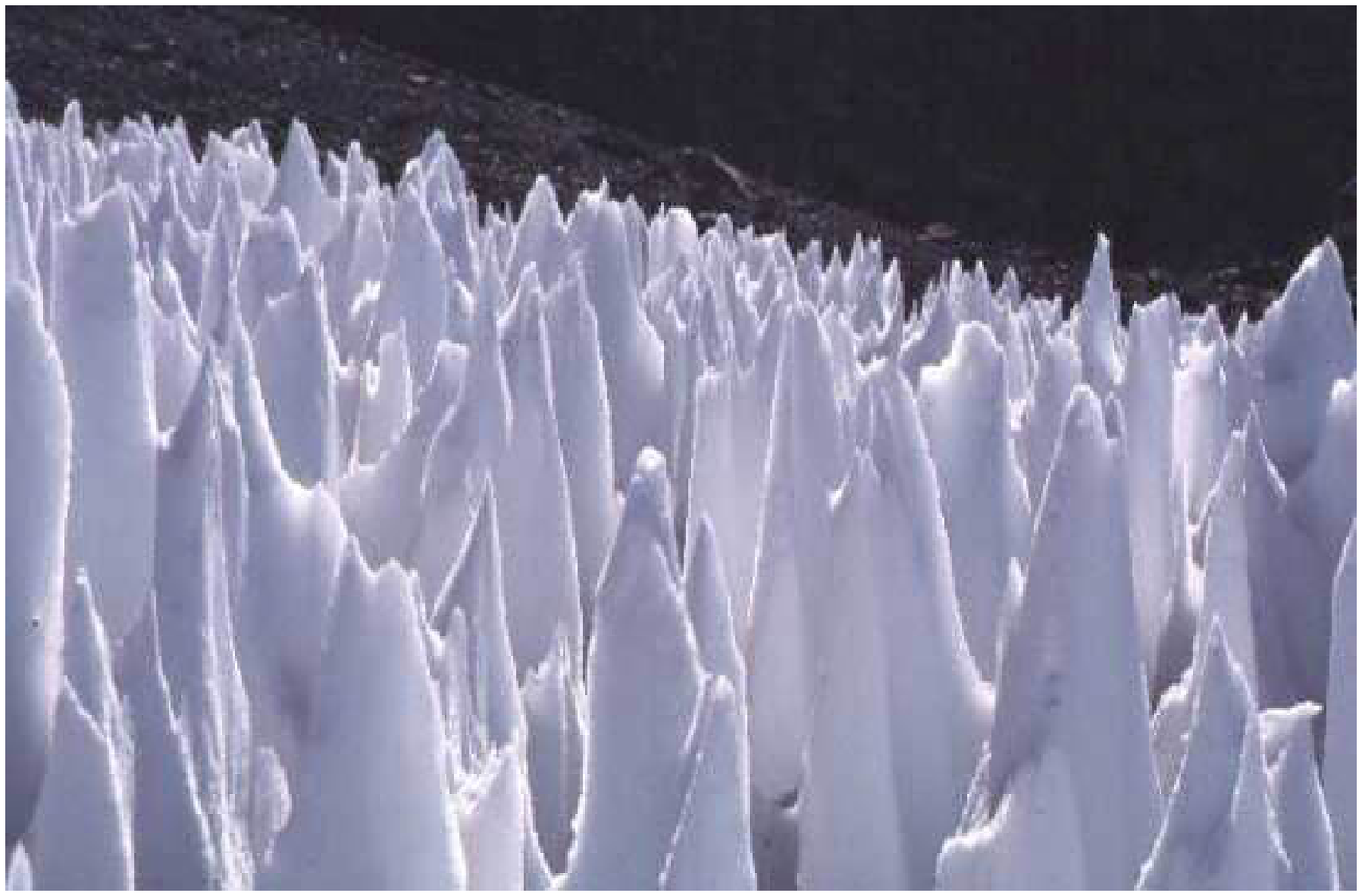}\includegraphics[height=5cm]{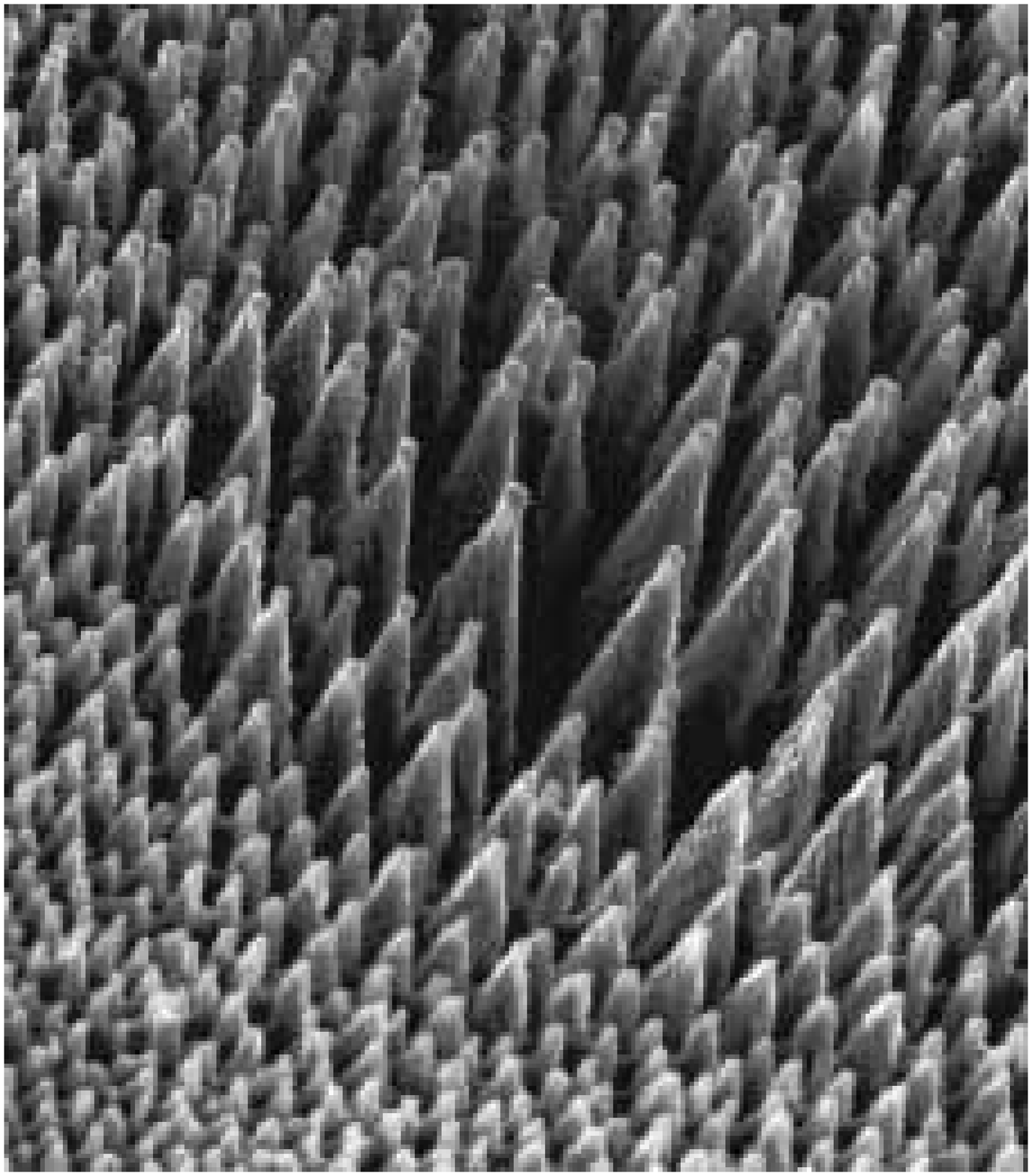}\includegraphics[height=5cm]{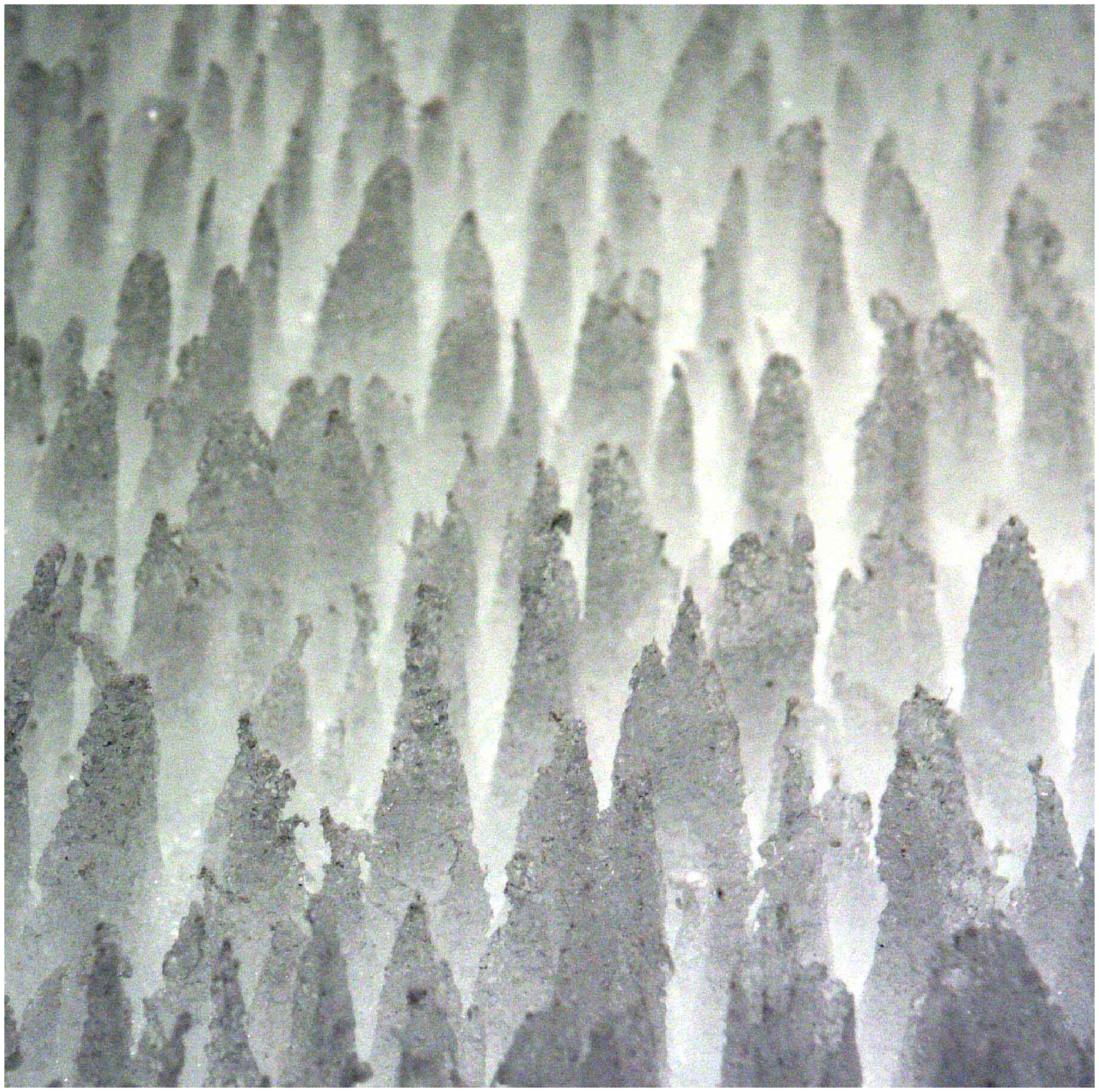}
\caption{
  Images of ablation spikes. (a) Penitentes in the Andes,
  approximately 2 m high\cite{corrip02}. (b) Silicon spikes created
  through laser ablation\cite{her98}. (c) Laboratory penitentes
  approximately 5 cm high. }
\label{fig1} 
\end{figure*}

Penitentes 1-4 m high are found on glaciers in high mountain regions;
laser ablation of materials can produce similar structures 10-100
$\mu$m high (Fig.\ \ref{fig1}). These structures are initiated by the
same underlying physics: ablation caused by direct and reflected
radiation. When radiation illuminates a surface, small surface
depressions receive more reflected light than high points, leading to
greater ablation in troughs and surface
instability\cite{rhodes87,better01,usosk99}.  Penitente formation
alters glacial energy balances and therefore affects local water runoff
and flooding, feedback in climate dynamics, and paleo-climatic
reconstruction\cite{kotly74,corrip02}. In materials science, surface
micropatterning via laser ablation can produce problematic surface
roughness\cite{folty94,miller98}, but also has several proposed
applications, including improved solar cells and electron-field
emitters\cite{her98,lo03}.

Here we report the first laboratory generation of snow penitentes 1-5
cm high (Fig.\ \ref{fig1}). We characterize lab penitente formation
and measure penitente coarsening. The lab setting allows controlled
environmental conditions, and we can measure the evolution of lab
penitentes visually in real time. Our experiments provide a model
system for studying ablation morphologies.

Penitentes occur on high-altitude snowfields exposed to intense
sunlight, particularly in South America\cite{llibout54}. Expeditions
studying penitentes have invoked the importance of
sunlight\cite{kotly74,corrip02,llibout54,troll42} and cold, dry
conditions (dew point below 0$^o$C), where melting is disfavored and
ablation proceeds by sublimation\cite{corrip02,llibout54}.
Concentrated reflections and protection from wind increase penitente
growth; higher temperature and humidity in the troughs of penitentes
allows melting in the troughs\cite{llibout54}. This effect accelerates
penitente growth, because sublimation of ice at 0$^o$C requires 7.8
times more energy than melting an equivalent volume, while the
sublimating surface is evaporatively cooled, further lowering the
surface temperature. Thus initiation of penitente growth requires a
low dew point, but melting can occur during later height growth.

Centimeter-scale penitentes similar in appearance to those we describe
have been observed in field work, where they are called
micropenitentes and are believed to be precursors to large Andean
penitentes\cite{corrip02,llibout54,llibout65}. Thus our experiments
may mimic the initial stages of penitente development on high-altitude
glaciers, and allow
further understanding of the conditions necessary for penitente
development.  Our observation contrasts with reports on the formation of
suncups, an ablation morphology found on lower-altitude glaciers and
snowfields with a typical initial wavelength of 5-10
cm\cite{rhodes87,better01}. Suncups typically form in melting snow,
where
the uniform temperature of melting means that no temperature gradients
along the surface are present.
%
%percolating meltwater eliminates temperature gradients along the
%surface. 
This effect may be responsible for the large difference in initial
feature size between micropenitentes and suncups.

Once penitentes are formed, feedback by multiple reflections and
evaporative cooling preserve their shape\cite{corrip02}.  Here we
study the initial stages of the instability when penitentes are most
vulnerable to disruption.  Simulations of an Andean glacier basin
suggest that penitentes lower the net ablation rate, preserving
glacier mass.  Global warming is expected to increase the minimum
elevation of penitente formation, accelerating the disappearance of
Andean glaciers and depleting water resources\cite{corrip02}.
Particulate pollution could futher accelerate Andean glacier
disappearance, because snow optical properties are significantly
altered by surface debris\cite{wiscom80,warren80,warren84}. Small dirt
particles adhere to the ablating surface and migrate toward its
peaks\cite{rhodes87}. A thin debris layer suppresses reflections and
disrupts penitente formation, while a thicker layer forms an
insulating sheath and promotes formation of different
structures\cite{rhodes87,better01}.

Laser ablation of surfaces can produce a variety of surface
patterns\cite{folty94}, including arrays of conical spikes that bear a
striking resemblance to penitentes. The enhanced light capture
due to multiple reflections on such micropatterned silicon surfaces
has led to the suggestion that they could form improved solar
cells\cite{her98}.  Conical spikes have been demonstrated in metals
and
semiconductors\cite{her98,sanch99,dolgaev01,kawak02,pedraz03,gyorg03},
polymers\cite{novis88,krajn93,hopp97,zeng99,wang03},
composites\cite{ono98} and ceramics\cite{folty94,jeong97,heitz97}. The
formation mechanism of ablation spikes shares common features with
penitente formation. The spikes tilt toward the illumination
direction\cite{her98,dolgaev01,krajn93} and multiple reflections are
believed to be important for their
growth\cite{usosk99,dolgaev01,pedraz03}. Other variables are also
important in the formation of microstructures by high-intensity laser
pulses, including light interference, hydrodynamics of melted surface
material, re-freezing of material at cone tips, surface impurities,
and chemical interactions with ambient gas.

\begin{figure}[b] 
  \centering \includegraphics[width=10cm]{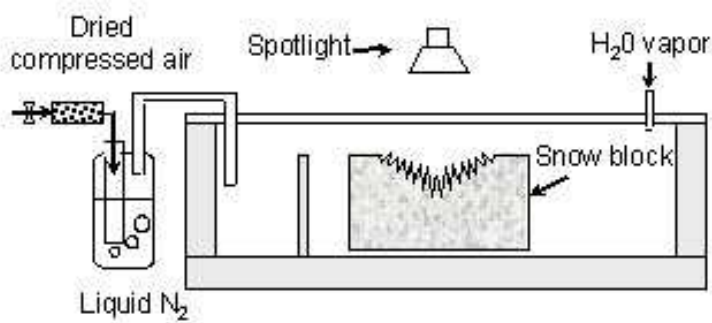}
\caption{
  Experimental apparatus. A horizontal freezer with Plexiglas cover
  serves as the environmental chamber. Temperature control is achieved
  by passing dried compressed air through liquid nitrogen and into the
  chamber. Humidity control is achieved by introducing water vapor.
  For further details, see \cite{suppl}.}
\label{fig2} 
\end{figure}

Our apparatus allows controlled air temperature down to -50$^o$C with
a relative humidity $<5$\% (Fig.\ \ref{fig2}).  Illumination is
provided by a flood lamp. We produce artificial snow with controlled
particle radius by freezing water droplets\cite{suppl}.  In our
experiments, penitente-like structures 1-5 cm high were produced after
a few hours of illumination. The direction of penitente growth is
always aligned with the incident angle of our radiation source as
observed in laser ablation and on glaciers.  Furthermore, at normal
incidence we were able etch the conical irradiation profile of the
lamp into the snow as shown in Figs.\ \ref{fig2} and \ref{fig4}.

Filtering the light or creating wind eliminates penitente formation in
our experiments. The absorption of light by snow is much higher for
near-infrared and infrared wavelengths than for other spectral
regions\cite{wiscom80}; therefore, the intensity of these wavelengths
controls the ablation rate. Selectively filtering out infrared
wavelengths inhibits penitente formation in our setup.  Similarly,
using a fan to provide a steady breeze of 2.5 m s$^{-1}$ over the snow
surface eliminates penitente production. The wind may prevent
temperature gradients along the surface needed for penitente formation
through a reduction in the thickness of the water vapor boundary
layer.

The optical extinction length determines the initial size structures
that form, except in the case of particle sizes comparable to the
extinction length.  We created artificial snow particles with radii
between 25 $\mu$m and 2.5 mm.  In the field, snow flakes age to become
quasi-spherical with age-dependent particle radii of 20 $\mu$m--1.5
mm, and snow optical properties change with
age\cite{wiscom80,warren80}.  The size of the smallest ablation
structures is limited by the optical extinction length: points on the
snow within one extinction length experience the same light intensity
and therefore ablate at the same rate\cite{better01}. The optical
extinction length is proportional to $r/v$, the particle radius
divided by the ice volume fraction in the snow\cite{wiscom80}. For our
dry, fine-grained snow, we estimate that the extinction length is
0.5--1 mm for particle radii 25--500 $\mu$m.  The snow density
increases with particle size.  Therefore as the particle radius
increases the volume fraction also increases and the extinction length
is approximately constant.  An extinction length of 1 mm is consistent
with our observation of an initial feature wavelength of approximately
3 mm, which we observe for all but very large ($r$=2.5 mm) particles.
For 2.5-mm radius particles, the formation of initial structures
requires a much longer time, and structures initially appear on the
scale of the particle size (5-mm wavelength).

\begin{figure}[t] 
  \centering 
\includegraphics[width=6cm]{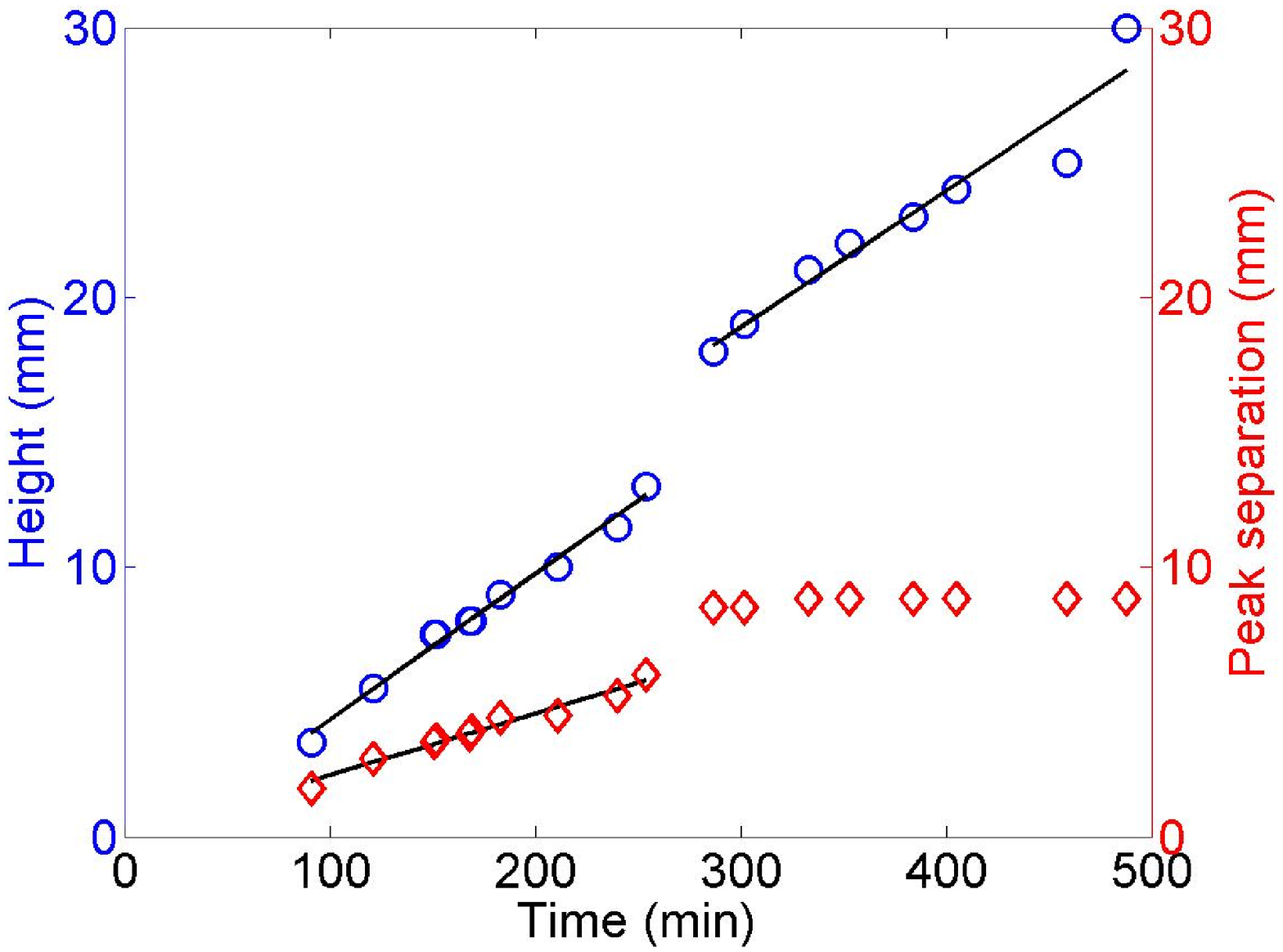}
\includegraphics[width=6cm]{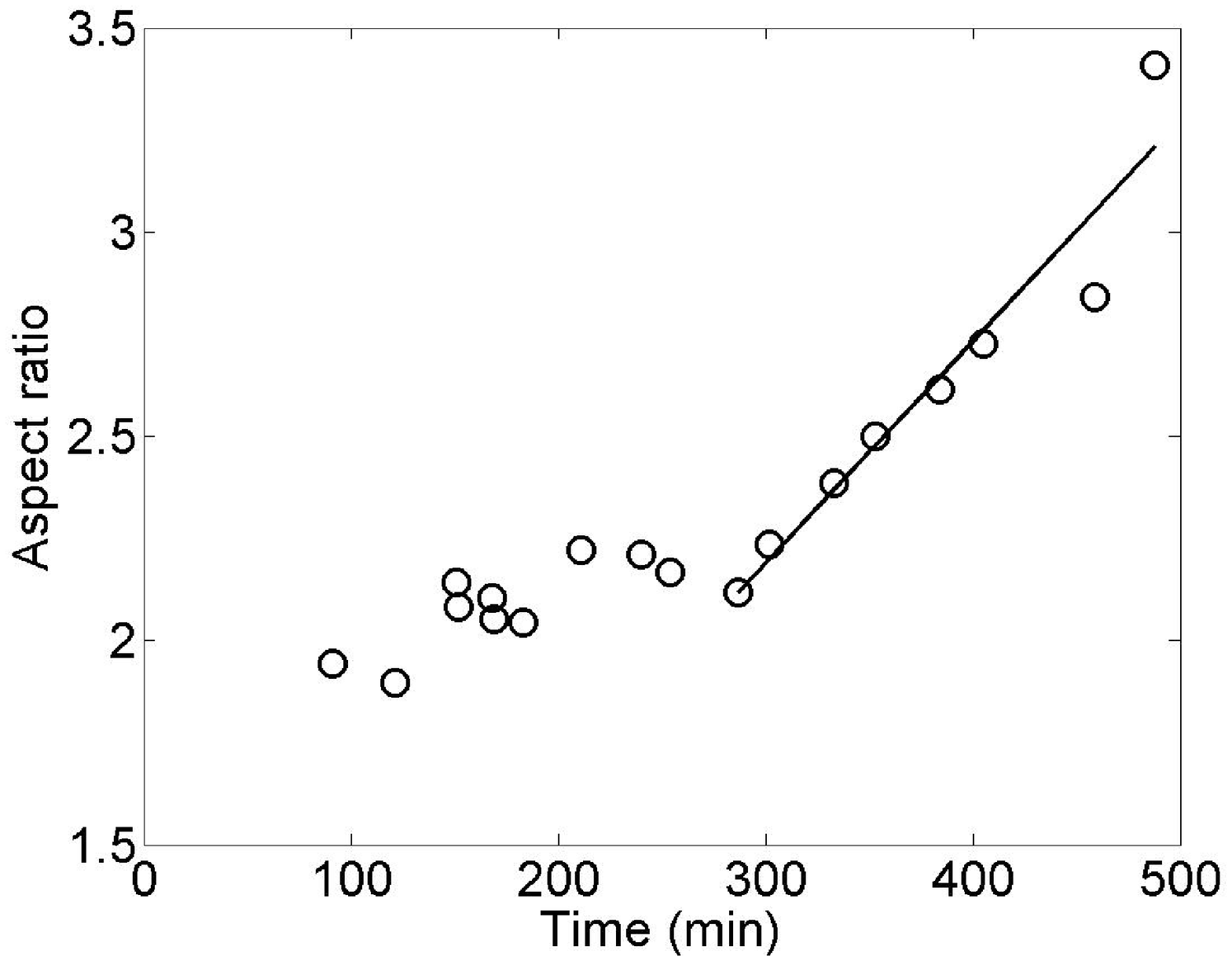}
\caption{
  Coarsening data. Top, measurements of penitente height (circles) and
  peak separation (diamonds) versus time. Bottom, penitente aspect
  ratio (height/peak separation) versus time. After an initial
  coarsening phase $\sim$300 minutes in length, the dynamics
  transition from sublimation throughout the structures to melting at
  the troughs and coarsening ceases.  In this experiment the initial
  particle radius was approximately 50 $\mu$m, infrared radiation
  intensity approximately 300 W m$^{-2}$, temperature -15$^o$C and
  relative humidity 10\%.}

\label{fig3} 
\end{figure}

Formation of penitentes does not require granular snow; we also
studied structure formation using a solid ice block. Similar
penitentes formed, but the initial development of structure required
approximately two times longer than for experiments with snow particles
(at the same temperature and relative humidity).  An opaque, frosty
layer formed on the ice surface before the formation of ablation
spikes. This layer may form when surface ice sublimates and the
resulting evaporative cooling of the surface causes a portion of the
water vapor to recondense. The surface roughness produced by the
recondensation may be important for the initial development of ice
surface structure.

After the initial instability forms surface structure, penitente
growth enters a nonlinear regime. The penitentes increase in size by
coarsening (Fig.\ \ref{fig3}), as predicted in a previous model of
penitente growth\cite{better01}. In our experiments, penitente height
and peak separation increase linearly, while maintaining an
approximately constant aspect ratio of 2. This value is comparable to
aspect ratios of 1.5-1.7 measured for Andean penitentes 1.35 m
high\cite{corrip99}.

In many experiments, we observe that coarsening stops after a certain
time. This dynamical transition occurs near $t$=300 minutes for the
experiment shown in Fig.\ 3.  After the transition, penitente height
continues to increase while the peak separation remains constant. In
addition, we observe a rounded shape at penitente tips after the
transition (Fig.\ \ref{fig4}). We hypothesized that the end of
coarsening behavior is associated with a transition to melting in
penitente troughs. Because larger penitentes capture a greater
fraction of the illuminating radiation, the reflected intensity in
penitente troughs increases with time during coarsening.  This
hypothesis was confirmed by experiments with a temperature-sensitive
dye added to the snow\cite{suppl}. In these experiments, the surface
layer was blue, corresponding to a warmer temperature than the pink
snow below the penitentes.  The transition to height increase without
coarsening was accompanied by a transition to melting: we observed
both downward seepage of warmer (blue) material and the elimination of
observable temperature gradients along the surface. This behavior is
expected for uniformly melting snow.  The transition to melting did
not occur for experiments at colder temperatures.

\begin{figure}[t] 
  \centering \includegraphics[width=6cm]{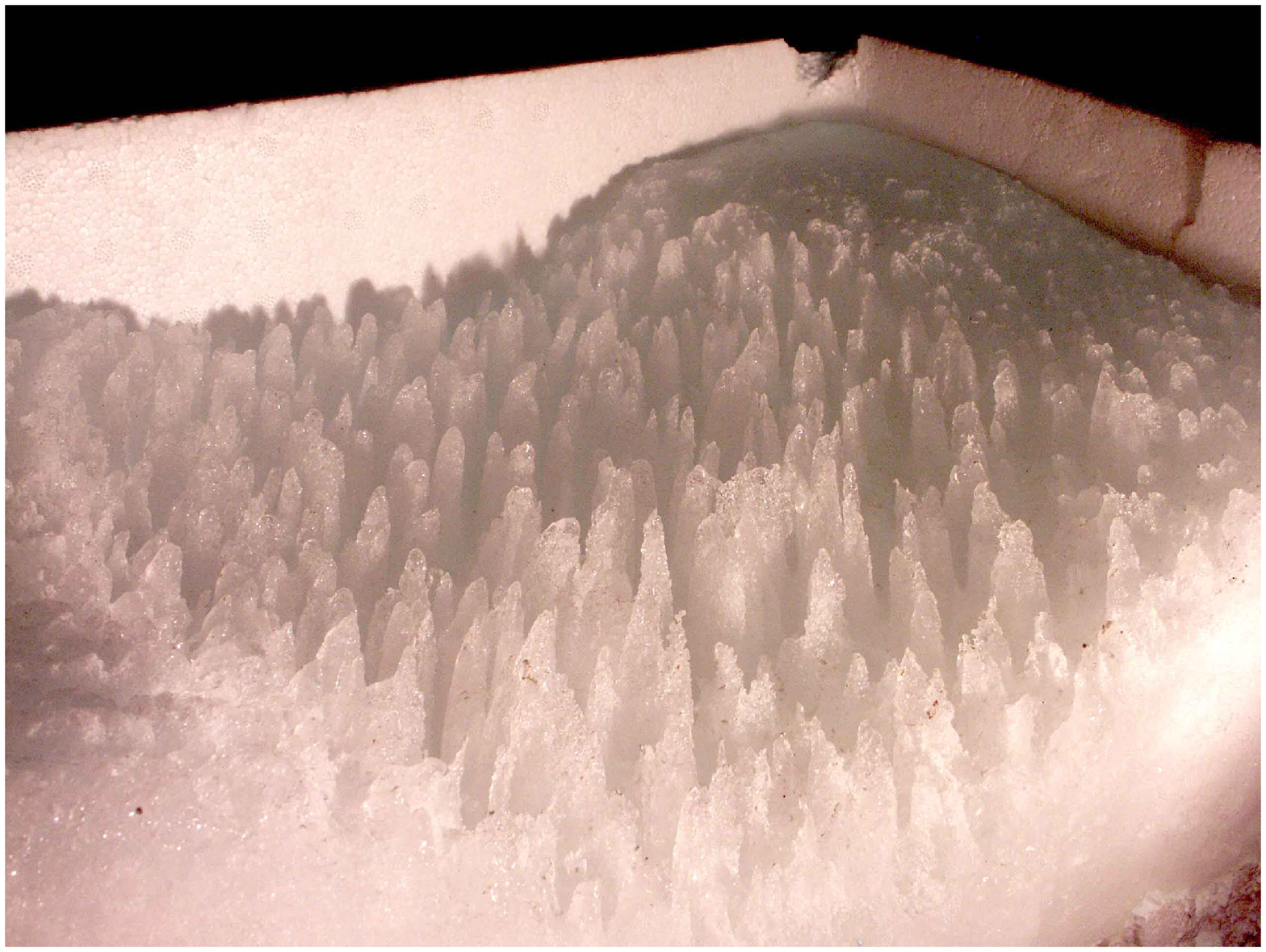}
\includegraphics[width=6cm]{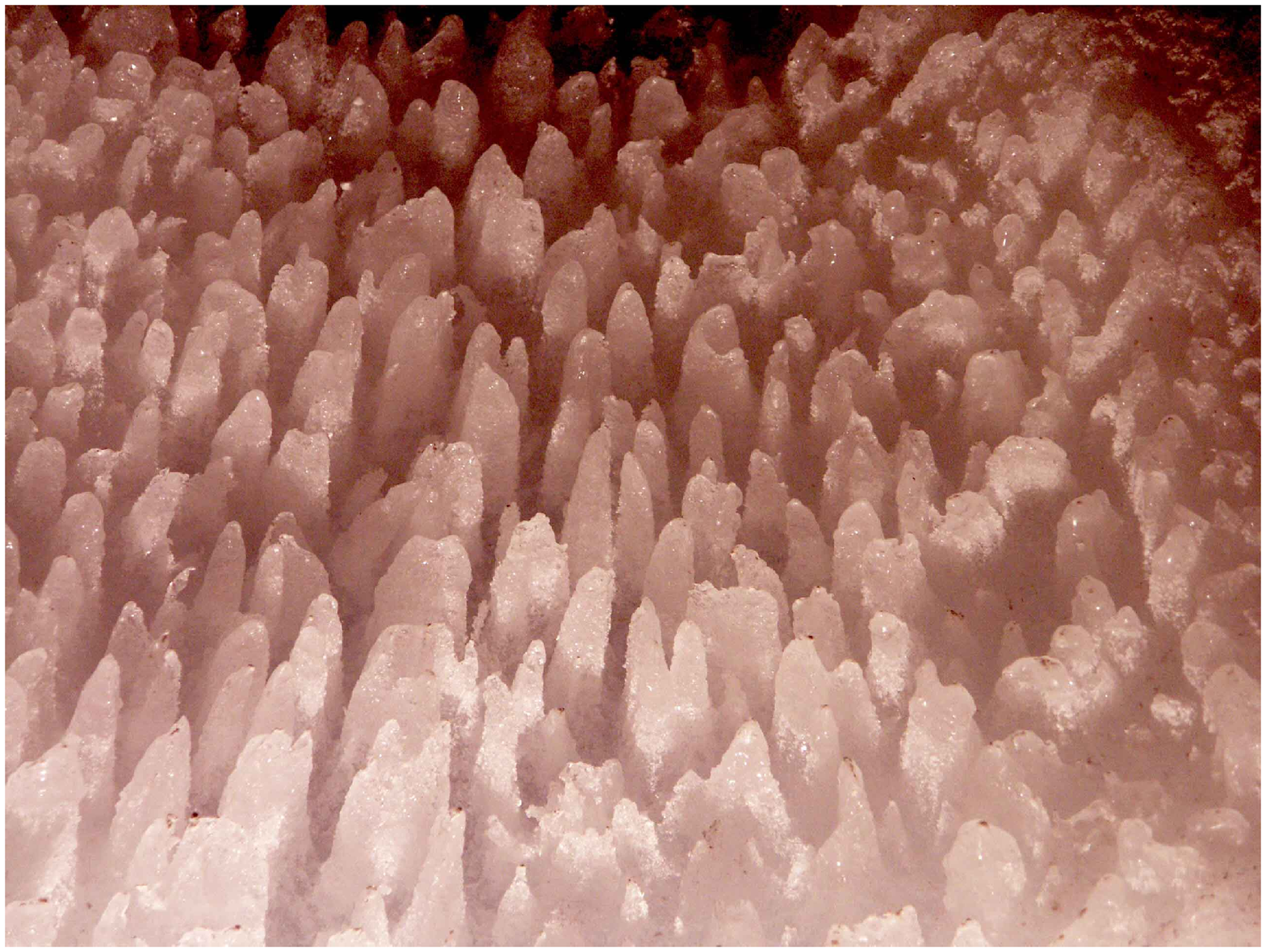}
\caption{
  Images of penitentes. Left, a snow block showing larger penitentes
  in the center where the light illumination is most intense. The
  curved shape of the surface is caused by the conical irradiation
  profile of the flood lamp. The snow block is 80 by 50 cm and the
  largest, central penitentes are 6 cm high.  Right, field of
  penitentes after the transition to melting in the troughs has
  occurred. The penitente aspect ratio increases and the tips become
  more rounded. The largest penitentes are 3 cm high.  }
\label{fig4} 
\end{figure}

Humidity changes have little effect on structure formation in our
experiments, while changes in temperature significantly alter the
growth of penitentes. For the conditions of Fig.\ 3, addition of water
vapor to the chamber to increase the relative humidity up to 70\% does
not significantly change penitente growth. However, temperature
changes significantly affect the process. Under melting conditions
(above -4$^o$C), the snow ablates and meltwater accumulates beneath
the snow without structure formation. Between -10 and -20$^o$C
penitentes appear within two hours of irradiation and coarsen as in
Fig.\ 3. Below -35$^o$C we could not produce penitentes after ten
hours of irradiation.  The strong temperature dependence suggests that
glacial penitentes are sensitive to temperature increases induced by
climate change: as global temperatures increase, regions where
penitentes currently form may become too warm for their growth.

The large change in vapor pressure with temperature is likely
responsible for the variations we observe. The sublimation rate is
proportional to the difference between the saturated vapor pressure
near the surface and the ambient partial pressure far from the
surface. The vapor pressure of water decreases
approximately exponentially between 0 and -35$^o$C, with a decay
constant $\sim$10.5$^o$\cite{crchandbook}.
Presumably, at -35$^o$C the low saturated vapor
pressure (20 Pa) impedes sublimation because the air near the ice
surface is saturated with water vapor.
The thickness of the boundary layer of water vapor near the ice
surface also affects the ablation rate.

To investigate the effects of surface debris on penitente formation,
we spread carbon black powder over half the surface of the snow block.
The powder layer thickness was comparable to the snow particle size,
the limit of the thin-dirt regime of ref.\ \cite{better01}. In the
debris-covered region, we observed earlier formation of penitentes
with larger peak separations than in the clean-snow region. In
addition the particulate material accumulated at the tips of the
penitentes. It is known that small dirt particles typically adhere to
an ablating surface and migrate toward its peaks\cite{rhodes87}. These
observations are consistent with a model of how dirt influences
ablation-structure formation\cite{better01}.

The result of our dirty-snow experiment
has surprising implications for the effects of particulate pollution
on penitente formation. On a relatively flat glacier surface, a thin
debris layer accelerates glacier ablation\cite{warren84}. However,
simulated glaciers covered with penitentes ablate more slowly than
flat ones,
because of shadowing and increased cooling by wind\cite{corrip02}.
Large fractions of a penitente-covered glacier are in shadow as the
sun moves throughout the day and the solar zenith angle changes from
day to day. Although penitentes absorb a slightly larger fraction of
the incident radiation than does a flat surface, the shadowing effect
dominates the increased absorption. In addition, a penitente-covered
glacier has larger surface area and is more effectively cooled by a
light wind blowing over the surface.
Thus, by accelerating penitente formation and producing larger
penitentes, small amounts of particulate matter could help preserve
Andean glaciers.

\begin{acknowledgments}
  We thank M.\ Ben Amar and J.\ Corripio for useful discussions. MDB
  acknowledges funding from the Alfred P. Sloan foundation.
\end{acknowledgments}  

%\bibliography{snow}

\end{document}